\documentstyle[preprint,aps]{revtex}
\tightenlines
\begin{document}

\title{Skyrmionic excitons}

\draft

\author{T. Portengen, J. R. Chapman, V. Nikos Nicopoulos, and N. F. Johnson}
\address{Department of Physics, University of Oxford,
Parks Road, Oxford OX1 3PU, United Kingdom}

\date{July 2, 1997}

\maketitle

\begin{abstract}
 We investigate the properties of a Skyrmionic exciton
 consisting of a negatively charged Skyrmion bound to 
 a mobile valence hole. A variational wave function is 
 constructed which has the generalized total momentum 
 ${\bf P}$ as a good quantum number. It is shown that the 
 Skyrmionic exciton can have a larger binding energy than 
 an ordinary magnetoexciton and should therefore dominate 
 the photoluminescence spectrum in high-mobility quantum 
 wells and heterojunctions where the electron-hole 
 separation exceeds a critical value. The dispersion 
 relation for the Skyrmionic exciton is discussed.
\end{abstract}


 Excitons play a crucial role in our understanding of the optical 
 properties of direct-gap semiconductors. The properties of excitons 
 in systems of reduced dimensionality in a high magnetic field were 
 first investigated by Lerner and Lozovik \cite{Lerner} many years
 ago, and have been the subject of numerous studies \cite{Korolev} 
 since. Recently, the existence of unusual charged excitations, known 
 as Skyrmions, was predicted theoretically \cite{Sondhi}, and confirmed
 experimentally \cite{Barrett,Schmeller,Aifer}, in quantum Hall 
 systems with filling factor near $\nu = 1$. A Skyrmion is a 
 topological twist in the spin density of the two-dimensional (2D) 
 electron gas. According to the sense of the twist, the Skyrmion 
 carries electrical charge $-e$ or $+e$. The size of the Skyrmion 
 is determined by a competition between the Zeeman energy and the 
 electron-electron exchange interaction. The relative strength of
 these two competing effects is conveniently parameterized by the
 ratio $g = \frac{1}{2}|g_{e}|\mu_{B}B/(e^{2}/\epsilon \ell)$, where
 $g_{e}$ is the Land\'{e} $g$ factor ($g_{e} = -0.44$ in GaAs), 
 $\ell = \sqrt{\hbar c/e B}$, and $\epsilon$ is the background 
 dielectric constant. The Zeeman energy favours small Skyrmions, 
 while the exchange interaction prefers large ones \cite{Fertig}. 
 Ordinary spin $\frac{1}{2}$ quasielectrons (quasiholes) can be 
 regarded as negatively (positively) charged Skyrmions of zero size. 
 
 In this paper we show that, just as an electron and a hole may bind
 to form an exciton, a negatively charged Skyrmion and a hole may bind 
 to form a {\em Skyrmionic} exciton. The Skyrmionic exciton can replace
 the ordinary magnetoexciton as the lowest energy excitation probed
 by a photoluminescence (PL) experiment \cite{Goldberg,Turberfield}
 at $\nu = 1$. Previously \cite{Portengen} 
 we considered the optical recombination of a 
 Skyrmion with a hole {\em localized} near a minimum in the disorder
 potential. The Skyrmionic exciton considered here is different in 
 that the Skyrmion and the mobile hole can move together throughout 
 the plane perpendicular to the magnetic field. The Skyrmionic 
 exciton hence has its own dispersion relation in terms of a quantum 
 number ${\bf P}$ which plays the role of the total transverse 
 momentum in a magnetic field. The key element in our approach is 
 the construction of a variational wave function with a definite value
 of ${\bf P}$, which produces an equal probability of finding the 
 Skyrmionic exciton anywhere in the plane perpendicular to the 
 magnetic field. This allows us to compute  
 the energy and the spin of the Skyrmionic exciton as a function of 
 ${\bf P}$.\@ We focus in particular on the state
 with ${\bf P} = 0$ which dominates the optical recombination spectrum 
 at $\nu = 1$. 

 We study a model system consisting of a 2D electron gas lying in the 
 $x$-$y$ plane, interacting with a valence hole confined to a plane at 
 a distance $d$ from the electron gas, in a strong magnetic field along 
 the $z$ direction. The confinement of the hole along the $z$ direction 
 occurs naturally in an asymmetrically doped quantum well \cite{Goldberg},
 and may be achieved in a single heterojunction \cite{Turberfield} by the
 application of a bias voltage. The strong magnetic field restricts 
 the electrons and hole to the lowest Landau level. The electrons 
 can have spin up or down, while the hole can have 
  $m_{j} = \pm \frac{3}{2}$. The label $m_{j}$ plays no role other 
 than to select the polarization of the luminescence \cite{Cooper}.
  Hence we drop this label and
 leave it understood that for a right-circularly polarized (RCP) 
 transition the hole must have $m_{j} = -\frac{3}{2}$, while for a
 left-circularly polarized (LCP) transition it must have 
 $m_{j} = +\frac{3}{2}$. The Hamiltonian is 
\begin{eqnarray}
 H & = & \varepsilon_{h} \sum_{m} h^{\dagger}_{m} h_{m} + 
      \frac{1}{2} g_{e}\mu_{B} B  \sum_{m} (
       e^{\dagger}_{m\,\uparrow} e_{m\,\uparrow} -
       e^{\dagger}_{m\,\downarrow} e_{m\,\downarrow} )
  \nonumber \\ & & \mbox{} + 
   \frac{1}{2}  \sum_{\sigma \sigma'} \sum_{m m' m'' m'''}
 V^{ee}_{m m' m'' m'''} e^{\dagger}_{m\,\sigma}
    e^{\dagger}_{m'\,\sigma'} e_{m''\,\sigma'} e_{m'''\,\sigma} 
  \nonumber \\ & & \mbox{} +
   \sum_{\sigma}  \sum_{m m' m'' m'''}
      V^{eh}_{m m' m'' m'''} e^{\dagger}_{m\,\sigma} 
        h^{\dagger}_{m'} h_{m''} e_{m'''\,\sigma} .
\end{eqnarray}
 Here $e^{\dagger}_{m\,\sigma}$ creates an electron with
 spin $\sigma$ ($\sigma = \;\uparrow$ or $\downarrow$) in the 
 state $\phi_{m}({\bf r}) = (2^{m+1} \pi\, m!)^{-1/2} 
 r^{m} e^{-i m \phi} e^{-r^{2}/4}$ ($\ell = 1$),
 and $h^{\dagger}_{m}$ creates a hole in the state 
 $\phi^{*}_{m}({\bf r})$. The energy of the hole 
 ($\varepsilon_{h}$) is measured relative to the position of the 
 Fermi level at $\nu = 1$. A uniform neutralizing background 
 is added to the Hamiltonian in the usual way.

 The ground state of $H$ at $\nu = 1$ is the fully polarized
 state $|0\rangle = \prod_{m=0}^{\infty} e^{\dagger}_{m \uparrow}
 |{\rm vac} \rangle$, where $|{\rm vac}\rangle$ is the empty conduction
 band. The positive background ensures that the state $|0\rangle$ is 
 electrically neutral. The initial states probed by a PL
 experiment at $\nu = 1$ lie in the sector of the Hilbert space 
 with $n_{h} = 1$ and $n_{\downarrow}-n_{\uparrow} = 1$. 
 Here $n_{h}$ is the number of holes in the valence band, and
 $n_{\downarrow}$ and $n_{\uparrow}$ are the number of spin down
 electrons and spin up holes in the conduction band, respectively. 
 The initial states have one hole in the valence band, and a 
 negatively charged excitation in the conduction band. For a 
 magnetoexciton the negatively charged excitation is a spin down 
 electron. For a Skyrmionic exciton the negatively charged excitation 
 is a Skyrmion.

  A magnetoexciton has the generalized
 total momentum ${\bf P}$ as a good quantum number \cite{Gorkov}. 
  In general, conserved quantum numbers arise
 from the invariance of the system under a group of
 symmetry operations. The symmetry group associated with  
 ${\bf P}$ is the magnetic translation group \cite{Zak}. 
 Magnetic translations differ from ordinary translations by
 a gauge transformation, which compensates for the shift in the
 argument of the vector potential due to the translation. Unlike
 ordinary translations, magnetic translations do not commute
 when acting on charged excitations such as $e^{\dagger}_{m \sigma}$
 and $h^{\dagger}_{m}$. Hence ${\bf P}$ is not a good
 quantum number for a charged particle in a magnetic field.
 However, magnetic translations {\em do} commute when acting on
 neutral excitations such as $e^{\dagger}_{m\,\sigma}h^{\dagger}_{m'}$ 
 and $e^{\dagger}_{m\,\downarrow}e_{m'\,\uparrow}$.
 The initial states can therefore be chosen as eigenstates of ${\bf P}$.

 In the Hartree-Fock approximation, the wave function for a 
 Skyrmion and a hole localized at the origin is given by \cite{Fertig}
\begin{equation}
\label{eq:state_1}
 a^{\dagger}|0\rangle = \prod_{m=0}^{M-1} (-u_{m} 
 e^{\dagger}_{m+1\,\downarrow} e_{m\,\uparrow} + v_{m}) 
 e^{\dagger}_{0\,\downarrow} h^{\dagger}_{0} |0\rangle .
\end{equation}
 A variational wave function for a Skyrmionic exciton of 
 generalized momentum ${\bf P}$ is constructed according to
\begin{equation}
\label{eq:state_2}
 a^{\dagger}({\bf P})|0\rangle = 
   \int d^{2} {\bf R}\,e^{i {\bf P} \cdot {\bf R}}
  a^{\dagger}({\bf R}) |0\rangle ,  
\end{equation}
 with
\begin{equation}
\label{eq:state_3}
 a^{\dagger}({\bf R}) = M_{\bf R} a^{\dagger}M^{\dagger}_{\bf R} ,
\end{equation}
 where $M_{\bf R}$ is the magnetic translation operator. This 
 construction is analogous to the tight-binding method for 
 constructing extended Bloch states from localized atomic orbitals. 
 Since $M_{\bf R} M_{\bf R'} = M_{\bf R+R'}$ when acting on neutral
 excitations, 
\begin{equation} 
  M_{\bf R} a^{\dagger}({\bf P}) M^{\dagger}_{\bf R} = 
   e^{-i {\bf P} \cdot {\bf R}} a^{\dagger}({\bf P}) . 
\end{equation}
 This shows $a^{\dagger}({\bf P})|0\rangle$ is a state with a 
 definite value of ${\bf P}$. The quantum number ${\bf P}$ labels the 
 irreducible representations of the magnetic translation group within
 the sector of the Hilbert space containing the initial states. 
 Because $M_{\bf R}$ commutes with $H$, the states $a^{\dagger}
 ({\bf P})|0\rangle$ are orthogonal and uncoupled by $H$. The 
 energy of the Skyrmionic exciton may therefore be obtained by
 minimizing the expectation value of $H$ in the state $a^{\dagger}
 ({\bf P})|0\rangle$. The values of $u_{m}$ and $v_{m}$ that
 minimize the energy will depend on ${\bf P}$. For $u_{m} = 0$ and
 $v_{m} = 1$ we recover the wave function for a magnetoexciton
 in the lowest Landau level \cite{Lerner}.
    
 Using a variational wave function with $M = 14$ parameters, we have 
 minimized the energy of the state $a^\dagger({\bf P}=0)|0\rangle$, 
 for various values of the separation $d$ between the electron and 
 hole planes. Our results are shown in Fig.~\ref{fig:one}.
 Figure~\ref{fig:one}(a) shows the difference in 
 energy between the Skyrmionic exciton and the ordinary magnetoexciton.
 Figure~\ref{fig:one}(b) shows the corresponding difference in spin. 
 The size of the Skyrmionic exciton (i.e.\ the number of spin flips) 
 and its energy are controlled by three competing effects: the 
 electron-electron exchange interaction, the Zeeman energy, and the
 electron-hole interaction. Just as for a localized Skyrmion \cite{Fertig},
 the exchange interaction favours a large number of spin flips whilst the 
 Zeeman energy opposes this. The hole interacts with the 
 charge distribution of the Skyrmion, which becomes more diffuse as the
 Skyrmion size increases. The electron-hole interaction favours a 
 strongly peaked charge distribution and as such moving the hole towards
 the electron plane should decrease the number of spin flips.

 The three-way competition is manifest in Figure \ref{fig:one}.
 As the Zeeman energy is lowered, the Skyrmionic exciton becomes larger 
 in size and increasingly lower in energy than the magnetoexciton. When 
 the hole is moved towards the electron plane (decreasing $d$) the size
 of the Skyrmionic exciton drops and its energy moves closer to that of 
 the magnetoexciton. A similar reduction in the size of the Skyrmion
 due to its interaction with a charged particle has been found for 
 a Skyrmion bound to a localized valence hole \cite{Portengen} and
 for a Skyrmion bound to a charged impurity \cite{Brey}.
 For any given separation $d$ there is a threshold 
 to Skyrmionic exciton formation. When the hole is far removed from 
 the electron plane ($d = \infty$) the Skyrmionic exciton forms below 
 $g = 0.025$. As the hole moves towards the electron plane the threshold 
 drops to lower $g$. For the case of $d < \ell$ there is no Skyrmionic 
 exciton formation. This means that magnetoexciton states are the 
 appropriate initial states for systems in which the hole is close the 
 electron plane \cite{Cooper}. However for systems in which $d > \ell$ 
 the ${\bf P} = 0$ magnetoexciton is {\em not} the lowest 
 energy state, this now being the ${\bf P} = 0$ Skyrmionic exciton.
 As an example consider a high-mobility GaAs/Ga$_{1-x}$Al$_{x}$As 
 one-side doped quantum well, with an electron density of $n_{s} =  
 10^{11}$ ${\rm cm}^{-2}$ and a well width of $400$ \AA.\@ For this density, 
 the magnetic length at $\nu = 1$ is $\ell = 126$ \AA.\@ Based on our 
 calculations we expect that Skyrmionic excitons have a 
 lower energy than magnetoexcitons in such quantum wells.
 
 The qualitative behaviour of the ${\bf P} = 0$ Skyrmionic exciton 
 is unaffected by the number of variational parameters, $M$; more 
 parameters simply scale the results to larger numbers of spin flips 
 and a correspondingly larger energy difference. The maximum number 
 of parameters we can use for the ${\bf P} = 0$ state is $M = 14$ to 
 yield a tractable computing problem. We believe that this yields all 
 the qualitative physical features of the Skyrmionic exciton. For a 
 given number of variational parameters (fixed $M$) we find that the 
 energy and spin differences for the ${\bf P} = 0$ state are always 
 larger than for the localized state studied in Ref.~\onlinecite{Portengen}.
 By comparison with our earlier calculations for the localized state 
 using $M = 60$ parameters we therefore expect Skyrmionic excitons 
 with at least $3$ spin flips in a $400$ \mbox{\AA} wide quantum well
 with $n_{s} = 10^{11}$ ${\rm cm}^{-2}$.  

 We now compare the optical recombination spectrum of the 
 Skyrmionic exciton with the recombination spectrum of a 
 magnetoexciton against a filled $\nu = 1$ background. 
 The selection rules require the conservation of ${\bf P}$, 
 and a change in the $z$ 
 component of the total angular momentum by $+1$ ($-1$) unit 
 for a RCP (LCP) transition. At temperature $T = 0$ the initial
 state prior to recombination is the state with the lowest
 energy in the sector of initial states, i.e.\ the 
 ${\bf P} = 0$ Skyrmionic exciton. The selection rules 
 give a final state containing $|S_{z} \pm \frac{1}{2}|$
 spin waves with generalized total momentum ${\bf P} = 0$. For a 
 magnetoexciton ($S_{z} = -\frac{1}{2}$), the final state
 contains no spin waves in the case of RCP,
 and one spin wave in the case of LCP.\@ By
 Larmor's theorem, the energy of a ${\bf P} = 0$
 spin wave is $|g_{e}| \mu_{B} B$. Hence the 
 recombination spectrum of a magnetoexciton against a 
 $\nu = 1$ background consists, at $T = 0$ and in the 
 absence of disorder, of a sharp line in either polarization.
 For a Skyrmionic exciton ($S_{z} < -\frac{1}{2}$) the final
 state contains multiple spin waves. For $|S_{z} \pm \frac{1}{2}|
 > 1$ there is a continuum of final states, with energies
 between $|S_{z} \pm \frac{1}{2}| |g_{e}| \mu_{B}B$
 (corresponding to all spin waves having ${\bf P} = 0$),
  and $|S_{z} \pm \frac{1}{2}|  
 (|g_{e}|\mu_{B}B + \sqrt{\pi/2}\,e^{2}/\epsilon \ell)$ 
 (corresponding to widely separated 
 spin down electrons and spin up holes \cite{Kallin}). 
 Hence the recombination spectrum
 of a Skyrmionic exciton has an {\em intrinsic} width of 
 order $|S_{z} + \frac{1}{2}| \sqrt{\pi/2}\,e^{2}/\epsilon \ell$ 
 in RCP, and of order $|S_{z} - \frac{1}{2}| \sqrt{\pi/2}\, 
 e^{2}/\epsilon \ell$ in LCP.\@ Since one more spin wave is
 left after LCP recombination, the intrinsic width of the LCP 
 line exceeds that of the RCP line. A similar broadening of
 the line shape due to the formation of a spin texture has 
 been predicted in the electromagnetic absorption spectrum of 
 a Skyrmion bound to a charged impurity \cite{Brey}. Whereas
 an observation of the broadening of the absorption spectrum 
 would require the growth of specialized structures, the 
 broadening of the luminescence lines should be observable
 in standard high-mobility quantum wells and heterojunctions. 
 An additional {\em extrinsic} broadening of the luminescence 
 lines is due to finite temperature and the presence of disorder. 
 The extrinsic broadening of the recombination spectrum of a 
 magnetoexciton also yields an LCP line whose width exceeds
 that of the RCP line \cite{Cooper}, which can account for
 the difference in line width observed in narrow quantum 
 wells \cite{Plentz}. In wide quantum wells both the extrinsic
 and intrinsic mechanisms will contribute to the broadening
 of the luminescence lines. The difference in line width between
 the LCP and RCP lines should persist to low temperatures and
 high mobilities in wide wells, where the intrinsic mechanism
 dominates, but not in narrow wells, where only the extrinsic
 mechanism operates.
 
 The detailed PL spectrum is difficult to calculate owing 
 to the complicated final state after recombination of a 
 Skyrmionic exciton. However, the average PL energy can be
 found directly from the initial state $|i\rangle$ 
 using \cite{Apalkov} 
\begin{equation}
 \langle \omega \rangle = 
 \frac{\langle i|L^{\dagger}[L,H]|i\rangle}
      {\langle i|L^{\dagger}L|i\rangle} ,
\end{equation}
 where $L$ is the luminescence operator. Previously \cite{Portengen}
 we calculated the average PL energy for a disordered system 
 using the initial state $|i\rangle = a^{\dagger}|0\rangle$. 
 We have repeated the calculation for a disorderless system using 
 the initial state $|i\rangle = a^{\dagger}({\bf P}=0)|0\rangle$.
 As before, the $g$ dependence of the red shift of the LCP line 
 at $\nu = 1$ provides means to distinguish between the optical 
 recombination of a Skyrmionic exciton and the recombination
 of a magnetoexciton against a filled $\nu = 1$ background.
 For a Skyrmionic exciton the red shift increases with $g$,
 while for a magnetoexciton the red shift is independent of
 $g$. Using a wave function with $M = 14$ parameters, we
 find the red shift increases by 0.03 $e^{2}/\epsilon \ell$ 
 when $g$ varies from $0.025$ to $0.02$, with $d = 3 \ell$. 
 In practice, the $g$ factor may be varied by 
 tilting the magnetic field \cite{Schmeller} or by applying 
 hydrostatic pressure \cite{Maude}. In a tilted-field 
 experiment, the deformation of the hole $z$ wave function 
 by the parallel magnetic field ($B_{\parallel}$) leads 
 to an additional angle dependence of the luminescence 
 energy. The red shift remains unaffected by $B_{\parallel}$,
 provided the hole confinement potential is the same on both
 sides of $\nu = 1$. This is the case for an asymmetrically
 doped quantum well \cite{Goldberg}. In a single heterojunction,
 the hole is confined near the electron plane on the low-$B$ 
 side of $\nu = 1$, but unconfined on the high-$B$ side. 
 The disparity in the confinement potential causes a decrease 
 of the red shift with tilting angle {\em opposing} the Skyrmion 
 signature. This problem may be overcome by applying a gate 
 voltage to the heterojunction, which confines the hole along
 $z$ on both sides of $\nu = 1$.
 
 Thus far our discussion has focused on the ${\bf P} = 0$ state,
 which is the state that determines the PL spectrum at $T = 0$
 in the absence of disorder. States with ${\bf P} \neq 0$ can
 also be probed, for example using PL with a grating on the
 sample, or by means of resonant light scattering.
 In the remainder of the paper we derive two exact properties 
 of the ${\bf P} \neq 0$ Skyrmionic exciton: its electric dipole 
 moment and the asymptotic behaviour of the dispersion in the 
 high $|{\bf P}|$ limit. We also describe features of the dispersion
 relation obtained from our variational state $a^{\dagger}({\bf P})
 |0\rangle$.

 The electric dipole moment of the Skyrmionic exciton is given by
\begin{equation}
\label{eq:dipole}
     {\bf d} = \frac{c}{B^{2}} {\bf P} \times {\bf B} .
\end{equation}
 Accordingly the electric dipole moment of the Skyrmionic exciton
 is independent of its spin, and equal to the electric dipole
 moment of an ordinary magnetoexciton \cite{Lerner}. This can be 
 shown explicitly for the state given by 
 Eqs.~(\ref{eq:state_1})--(\ref{eq:state_3}),
 or more generally as follows. The Gor'kov momentum for a system 
 of $N$ charges in a magnetic field is given by
\begin{equation}
\label{eq:Gorkov}
\hat{\bf P} = \sum_{i=1}^{N} {\bf \Pi}_{i} + \frac{1}{c}
       {\bf B} \times {\bf d} ,
\end{equation}
 where ${\bf \Pi}_{i}=-i \nabla_{i}-\frac{q_{i}}{c}{\bf A}({\bf r}_{i})$,
 and ${\bf d} = \sum_{i=1}^{N} q_{i} {\bf r}_{i}$ is the electric 
 dipole moment. We now take the expectation 
 value of Eq.~(\ref{eq:Gorkov}) in an eigenstate of $\hat{\bf P}$. 
 For a state in the lowest Landau level the expectation 
 value of $\sum_{i=1}^{N} {\bf \Pi}_{i}$ is equal to zero. Hence
 ${\bf P} = \frac{1}{c} {\bf B} \times {\bf d}$, where ${\bf P}$ is 
 the eigenvalue of $\hat{\bf P}$. Taking the cross product with 
 ${\bf B}$ yields Eq.~(\ref{eq:dipole}).  
 Equation~(\ref{eq:dipole}) is a general property of a Skyrmionic 
 exciton in the lowest Landau level, which does not rely on the
 specific form of the wave function in 
 Eqs.~(\ref{eq:state_1})--(\ref{eq:state_3}).

 Using Eq.~(\ref{eq:dipole}) we can establish the exact 
 asymptotic form of the dispersion relation of the Skyrmionic exciton 
 in the high ${\bf P}$ limit. Suppose we were given the exact ground 
 state $s^{\dagger}_{0}|0\rangle$ and energy $\varepsilon_{SK}$ for a 
 Skyrmion localized at the origin. Exact Skyrmion eigenstates have 
 been found for a hard-core model Hamiltonian \cite{MacDonald}. 
 We now use the state $a^{\dagger}|0\rangle = s^{\dagger}_{0}
 h^{\dagger}_{0}|0\rangle$ to construct the Skyrmionic exciton, 
 instead of Eq.~(\ref{eq:state_1}). The resulting state $a^{\dagger}
 ({\bf P})|0\rangle$ is an exact eigenstate of $H$ for $d = \infty$
 with energy $\varepsilon_{h} + \varepsilon_{SK}$. If 
 $s^{\dagger}_{0}|0\rangle$ is within the lowest Landau level, 
 the dipole moment of $a^{\dagger}({\bf P})|0\rangle$ is given by
 Eq.~(\ref{eq:dipole}). Because the Skyrmion has no {\em internal}
 dipole moment in its ground state, the separation between the 
 Skyrmion and the hole in the state $a^{\dagger}({\bf P})|0\rangle$
 is ${\bf r}_{0} = {\bf d}/e$. The electron-hole interaction breaks 
 the degeneracy of the states $a^{\dagger}({\bf P})|0\rangle$ with 
 respect to ${\bf P}$. For large $|{\bf P}|$ the energy shift is 
 $-e^{2}/(\epsilon |{\bf r}_{0}|)$. The exact asymptotic behaviour 
 of the dispersion law in the limit $|{\bf P}| \rightarrow \infty$ 
 is therefore given by $\varepsilon_{h} + \varepsilon_{SK} - e^{2}/
 (\epsilon |{\bf P}|\ell^{2})$.
 
 We have obtained a dispersion law for the Skyrmionic exciton by 
 minimizing the energy of the state $a^{\dagger}({\bf P})|0\rangle$ 
 with $M = 10$ parameters as a function of $|{\bf P}|$. The 
 dispersion is parabolic near ${\bf P} = 0$ with an effective 
 mass that differs by less than 7\% from the effective mass of 
 a magnetoexciton. The energy and spin differences between the 
 Skyrmionic exciton and the magnetoexciton become smaller as 
 $|{\bf P}|$ increases. While the variational state
 $a^{\dagger}({\bf P})|0\rangle$ gives reliable
 results for the dispersion at small ($< 1/\ell$) values of 
 $|{\bf P}|$, this state does not adequately describe the dispersion 
 at large ($ > 1/\ell$) values of $|{\bf P}|$. Above a certain
 value of $|{\bf P}|$ the magnetoexciton becomes the lowest energy
 state within the variational space, yielding a dispersion that 
 approaches 
 $\varepsilon_{h} + \frac{1}{2}|g_{e}|\mu_{B}B$ in the limit 
 $|{\bf P}| \rightarrow \infty$. However, from our previous 
 discussion we know that the exact dispersion must approach 
 $\varepsilon_{h} + \varepsilon_{SK}$. The incorrect behaviour 
 of the variational state $a^{\dagger}({\bf P})|0\rangle$ in 
 the high-$|{\bf P}|$ limit occurs because this state forces 
 the Skyrmion to develop an internal dipole moment as $|{\bf P}|$
 increases. The internal dipole moment---which must be absent in
 the exact state $a^{\dagger}({\bf P})|0\rangle$---prevents the
 formation of a spin texture at high $|{\bf P}|$. Increasing the
 number of parameters cannot suppress the internal dipole moment 
 and thus does not remedy the failure of our variational state 
 at high $|{\bf P}|$. The construction of a variational state that 
 {\em does} reproduce the correct asymptotic behaviour of the 
 dispersion in the high-$|{\bf P}|$ limit is left for future 
 research.  
  
 We thank Andrew Turberfield, Darren Leonard, Stuart Trugman and 
 Igor Lerner for helpful discussions. This work was supported by 
 EPSRC Grant No.\ GR/K 15619.

\begin{figure}
\caption{\label{fig:one}
 Difference in (a) energy and (b) spin between a 
 ${\bf P} = 0$ Skyrmionic exciton and a ${\bf P} = 0$ 
 magnetoexciton. The energy difference
 $\Delta E$ and spin difference $|\Delta S_{z}|$ are 
 plotted vs the reduced Zeeman energy $g =
 \frac{1}{2}|g_{e}|\mu_{B}B/(e^{2}/\epsilon \ell)$, 
 for various values of the separation $d$ between 
 the electron and hole planes. The number of variational 
 parameters is $M = 14$.} 
\end{figure}  

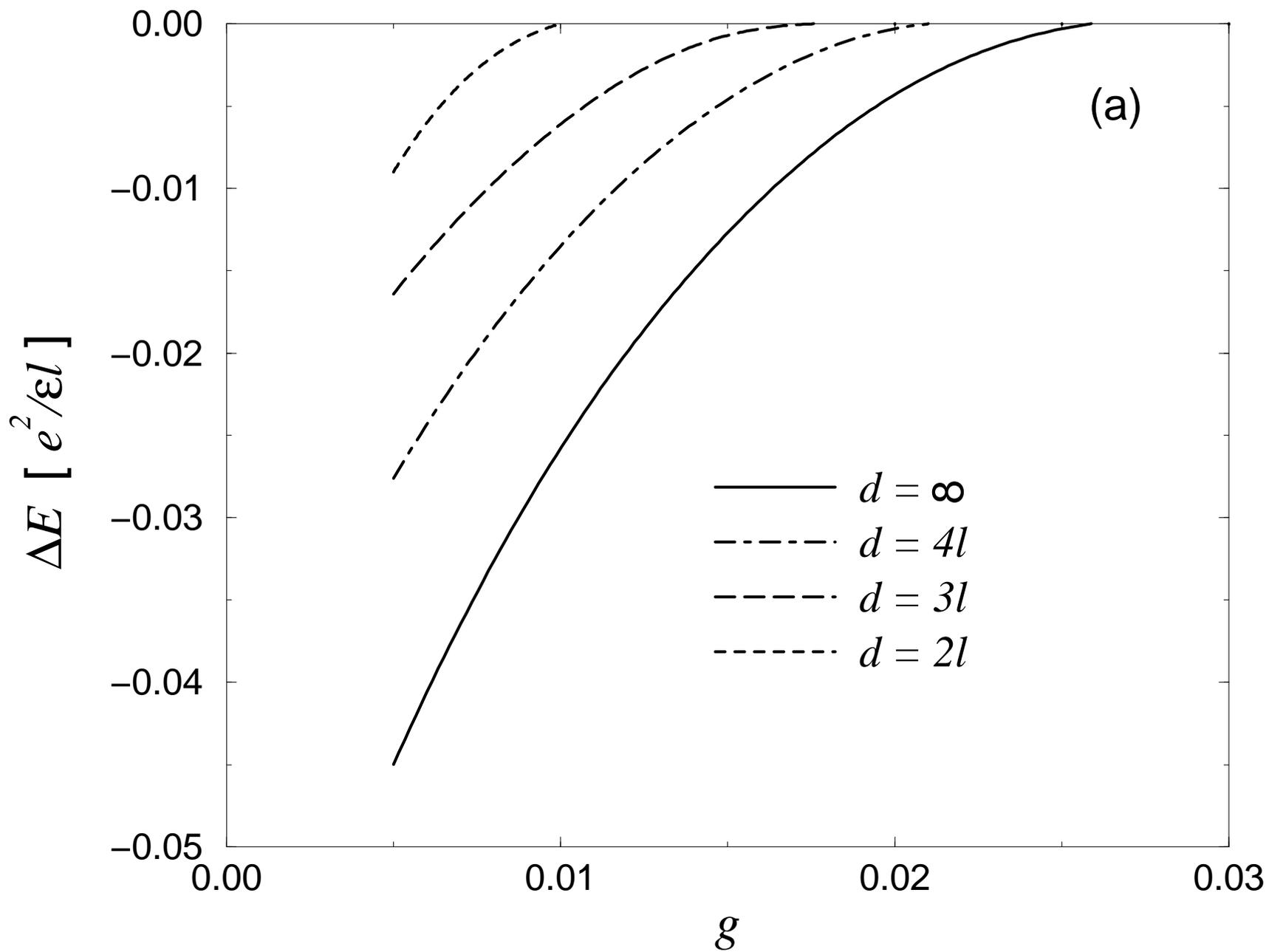
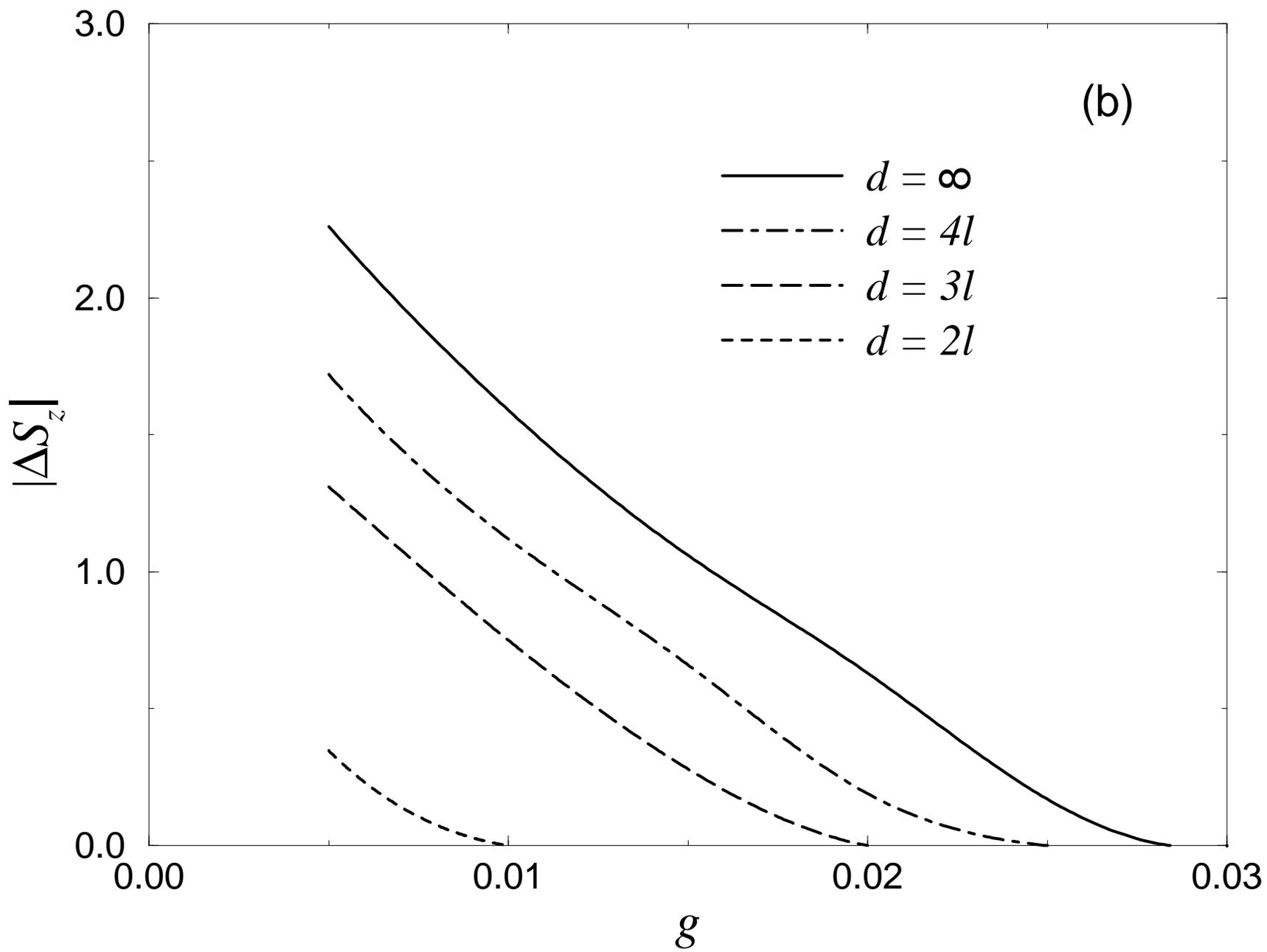
\end{document}